\documentclass{appolb}
\usepackage{graphicx}

\begin{document}

\title{Universalism versus particularism through ESS lenses\thanks{Presented at the 2nd Polish Seminar on Econo- and Sociophysics, Cracow, 21-22 April 2006}}
\author{Maria Nawojczyk\footnote{maria@list.pl}
\address{Faculty of Applied Social Sciences,
AGH University of Science and Technology,
Al. Mickiewicza 30, PL-30059 Cracow, Poland}
}

\maketitle

\begin{abstract}
The cultural variation of economic activity is wide and multidimensional. In my presentation I will refer to the analyses of the culture of capitalism provided by Alfons Trompenaars and Charles Hampden-Turner. According to them there are seven processes and related dilemmas which are important in analyzing the construction of a cultural system of economy. I will focus only on one of them, universalism versus particularism.
Using the database of Trompenaars and Hampden-Turner I will show how this dilemma was solved by managers from different European countries. That will be starting point for my analysis of universalism-particularism attitudes of respondents of European Social Survey (ESS). I will be particularly interested in verification of hypothesis  on the place of Poland on the mosaic of European cultures of capitalism.
\end{abstract}


\bigskip
\section{Introduction}

Globalization was initially presented as a process of ``simple'' homogenization of the market economy. At first its economic and political dimensions were underlined, but later, during last several years, a growing number of scholars have come to see the cultural dimension as an important phenomenon. They have concentrated here, however, on the Americanization of the world economy \cite{ref-3,ref-4}. On the other hand, it should be stressed that the students of macro-economy no longer question the heterogeneity of cultures of capitalism. Particularly after the fall of socialism, which resulted in the withering away of a viable alternative or competition to the capitalist system, discussion of the disadvantages of capitalism and its heterogeneity became even more vivid than before \cite{ref-5,ref-6,ref-7}.

At the same time, there has been a significant shift in theoretical reflection on these phenomena. In the social sciences, the strong influence of postmodernism results in a growing stress on the analysis of heterogeneity as such, on the studying of the specific rather than the general, and on interdisciplinary research. On the other hand, contemporary economic analysis, with its strong neo-institutional current \cite{ref-8}, draws on psychology and sociology.

Since the late 1980s the ``new economic sociology'' \cite{ref-9,ref-10} has been a very dynamic current attempting to unify sociological and economic traditions, and has proven itself to be a fruitful approach to research. In 1985 Mark Granovetter proposed the analysis of the embeddedness of economic processes in social structure, and his article became very influential, serving perhaps as a ``program manifesto'' for this current of thought. It opened the road to a number of interesting studies which brought a broader, more diverse view of the socio-economic realm.

These new studies went in several interesting directions, like, for instance a) attempts to bridge the gap between sociological and economic methodologies \cite{ref-11,ref-12}, b) an analysis of economic phenomena through the underlying social value systems \cite{ref-13,ref-14,ref-15}, and the most inspiring for me, c) an approach stressing the cultural background of economic activities \cite{ref-16,ref-17}. The latter has, in sociology, its classic antecedence in works of Max Weber \cite{ref-18,ref-19} and Thorstein Veblen \cite{ref-20}.

Theories of post-industrial society \cite{ref-21,ref-22} have also proven very insightful. I would particularly stress the concepts of ``flexibility'' and ``learning organizations'', as well as the concept of knowledge as a significant factor in social and economic development. These theories have marked a new way of looking at the economic processes taking place within a social structure and a culture \cite{ref-23,ref-24,ref-25,ref-26,ref-27}.

These sociological conceptions of economic organizations, as well as of culturally determined processes of doing business and managing, have found their application in the field of intercultural management \cite{ref-1,ref-2,ref-28,ref-29}. Scholars employing this method of analysis draw upon classics (including the modern classics) of sociology \cite{ref-18,ref-30,ref-31,ref-32,ref-33} and of social and cultural anthropology \cite{ref-34,ref-35}.

In the light of all these analyses, there is no doubt that today the concept of capitalism is, culturally speaking, heterogeneous \cite{ref-36,ref-37}, and that we have to do with at least three different patterns of capitalism: Anglo-Saxon, German and Asian. The German model, also called continental European, is internally heterogeneous as well and its homogenization is a painful process, very well visible in the problems with continuous integration of the European Union. All these patterns of cultures of capitalism have emerged in the centuries-long process of evolution, and have taken shape in a ``natural'' way, slowly building relatively coherent legislative and cultural systems. 

Thus the cultural variation of economic activity is wide and multidimensional. There are also different methodological approaches to the relations between culture and economy in which culture can be treated once as an independent variable and another time as a dependent variable.  On this simple binomial relation one can superimpose various levels of social analysis: macro-, mezzo- or micro. The thus established models could be further analyzed from sociological, economic, organizational, anthropological or mixed perspectives. Because of rising interest in these issues one can find examples of each of the above mentioned approaches. Like in every relatively new field of research, there are still more questions then answers. 
The text presented below is meant to participate in this discussion. So, within it, there are also more questions than answers.  To articulate these questions, I choose as an example the analyses of the culture of capitalism provided by Alfons Trompenaars and Charles Hampden-Turner \cite{ref-1,ref-2}.
 
\section{The model}
I will start my analysis with a short presentation of the assumptions in Trompenaars' and Hampden-Turner's model. Economic institutions are the basic level of their analysis, so their primary interest is in the culture of organization, as well as in intercultural management. Institutions are constituted by people, and the most important stratum for economic institutions is the managerial level. Therefore, their attitudes and their cultural values are the subject of research for Trompenaars and Hampden-Turner. Managers are influenced by the culture of the society within which they were socialized, but when shaping the organizations of their institutions and the relations between them, these managers also influence the culture of the whole society. So the authors are regularly switching between levels of analysis, from the attitudes of individual managers, to the roles operating within organizations, and to the ``culture of economy'' as such. The interactions between these levels are neither clear nor simple and they leave too much opportunity for unjustified generalizations. However, this way of reasoning, even if not very precise, can be attractive and stimulating. 

In order to understand how cultural values influence the choices made by people in the economic field, we must first distinguish which processes are, in this particular field, significant for the entire system, and then examine their cultural dimensions. There are seven such processes and related dilemmas which are important in analyzing the construction of a cultural system of economy.	The approach presented below was earlier applied by Charles Hampden-Turner and Alfons Trompenaars in their studies of international managers \cite{ref-2}.
\begin{description}
\item[Particularism vs. universalism.] The process of setting and using rules and standards to regulate economic activities on the macro-social level (economy), as well as on the micro-social level (individual enterprise), is very important. The efficiency of an economic system depends on the ability to set universal rules while still taking exceptions into account. Thus, some societies will more likely try to put everything under the control of general rules and others will be more willing to treat every situation as exceptional. The practice of economic behavior will reflect the particular solutions of this cultural dilemma between universalism and particularism.
\item[Analysis vs. synthesis.] Progress in any sphere depends on the ability to learn. This is based on the capacity to deconstruct and reconstruct, and reflects the cultural ability to think analytically or synthetically. As a result, economic processes can be organized into a series of detailed functions to be fulfilled (analysis) or into a single new scheme (synthesis).
\item[Individualism vs. collectivism.] Every economic activity is based on individual initiative, entrepreneurship and on one's drive to achieve certain goals. At the same time, the interests of groups, of the enterprise and of the whole economic system have to be taken into account. Thus, there is a constant cultural tension between individualism and collectivism. Depending on which of the two social values is dominating, the definitions of economic institutions and their functions will differ. This variation is particularly visible in human resources management (personal policy, benefits, evaluation etc.).
\item[Internal vs. external control.] In order to achieve certain goals the individual or group must make decisions. The question is whether they make decisions based on their own judgement, vision and responsibility or whether they take into account some external factors. The cultural dilemma is to what extent the majority of society is likely to be internally or externally controlled.
\item[Time sequential vs. synchronic.] Being successful in the market requires not only making the right decisions but also making them at the right time. The final effect can be achieved through sequence of steps, or by synchronizing many processes at the same time. This cultural approach to time is reflected in the organization of economic processes. 
\item[Ascribed status vs. achieved status.] Economic activity is realized by economic institutions that have a hierarchical structure. The rules of moving up in this structure are important for the development and stability of the organization. In cultures which stress achieved status, organizations are more flexible in economic activity but they have problems with management stability. In cultures that stress ascribed status, economic activity is less dynamic but management is more stabile. 
\item[Democracy vs. hierarchical structure.] The structure of an economic organization effects not only managers but also every other participant. There are democratic organizations, where more participants are involved in decision making processes and there are very hierarchical organizations based on giving and following orders.
\end{description}

Using a methodological approach characteristic for the economy, with a tendency toward model building and based on the dimensions presented above, we can construct two opposite ideal cultures of capitalism, obviously simplifying the actual variety of situations. One of these would be characterized by universalism, an analytic way of thinking, individualism, a linear approach to time, internal control, stress on achieved status and democracy. The American model of capitalism would be an example of this ideal type. The other ideal type would be characterized by particularism, a synthetic way of thinking, collectivism, a synchronic approach to time, external control, and stress on ascribed status and hierarchy. The Japanese model would be an example of this second ideal type. Between these poles there is a spectrum of various combinations of the dimensions presented above. These combinations have examples first and foremost in European models of capitalism. Using the concept of embeddedness taken from economic sociology Hampden-Turner and Trompenaars made detailed descriptions of these models looking within historical, political and cultural contexts for explanations of each particular cultural dilemma. Therefore, we find in their books references to ``national character'' as an explanation of individual behavior. This solution could be questioned for two reasons. First, it leads to the strengthening of national stereotypes. Second, in the history of nearly every society one can find examples where a set of completely contradictory values were nurtured. The mentioned kind of analysis can be considered as speculation or as stimulation for further discussion. I will treat it as a stimulating speculation.
I will focus only on one of them, universalism versus particularism. 

\section{The managers attitudes}
Using the database of Trompenaars and Hampden-Turner I will show how this dilemma was solved by managers from different European countries. Follow Hampden-Turner and Trompenaars I will present two stories which served them as a measures of particularism-universalism dilemma.

\begin{center}
\begin{tabular}{p{6cm}p{6cm}}
You are drinking beer with your friend who is responsible of safety at work during his duty hours. One employee become injured. You have to testify and you are the only witness.  What right has your friend to expect you to protect him?
&
You are manager of a big company's department. One of your employee with personal problems you are aware of is constantly delayed at work. What right has your employee to expect his fellow workers to cover him from you?
\\
\begin{enumerate}
\item[A.]
My friend has a definite right as a friend to expect me to testify to protect him.
\item[B.]
He has some right as a friend to expect me to testify to protect him.
\item[C.]
He has no right as a friend to expect me to testify to protect him.
\end{enumerate}
&
\begin{enumerate}
\item[A.]
He has a definite right to expect his fellow workers to cover him.
\item[B.]
He has some right to expect his fellow workers to cover him.
\item[C.]
He has no right  to expect his fellow workers to cover him.
\end{enumerate}
\end{tabular}
\end{center}

Percentage of C answers for both stories among the managers from different European countries: 
\begin{center}
\begin{tabular}{p{3cm}p{2cm}p{3cm}p{2cm}}
(first story)	&					&(second story)&\\
Germany 	& 90					&Germany 	& 94 \\
Sweden		& 89					&Sweden	& 91 \\
UK		& 82					&UK		& 84 \\
Belgium		& 67					&Belgium	& 57 \\
France		& 53					&France		& 43 \\
\end{tabular}
\end{center}

As above results shows, universalists are more common in Protestant cultures --- Germany, Sweden, UK. Predominantly Catholic cultures are more particularistic --- France, Belgium. These results are based only on analysis of one social stratum --- managers, and they are not representative sample of the whole society. But they are obviously very important social group and analyzing economic behavior we can treat managers as ``culture carriers''.

\section{The index}
That will be starting point for my analysis of universalism-particularism attitudes of respondents of European Social Survey \cite{ref-38}. In the data of ESS I have found three variables which can be treating as indicators of particularistic or universalistic attitudes of respondents. The distributions of these variables are presented below.
\begin{enumerate}
\item[I.]
If you want to make money, you can't always act honestly.\\
\begin{tabular}{l p{5cm}p{5cm}}
1.& Agree strongly&7.5\%\\
2.& Agree&27.8\%\\
3.& Neither agree or disagree&18.4\%\\
4.& Disagree&32.8\%\\
5.& Disagree strongly&13.5\%\\
\end{tabular}

\item[II.]
You should always obey law even if it means missing good opportunities.\\
\begin{tabular}{l p{5cm}p{5cm}}
1.& Agree strongly&16.3\%\\
2.& Agree&48.1\%\\
3.& Neither agree or disagree&22.8\%\\
4.& Disagree&11.0\%\\
5.& Disagree strongly&1.7\%\\
\end{tabular}

\item[III.]
Occasionally alright to ignore law and to do what you want.\\
\begin{tabular}{l p{5cm}p{5cm}}
1.& Agree strongly&2.3\%\\
2.& Agree&18.7\%\\
3.& Neither agree or disagree&21.8\%\\
4.& Disagree&42.2\%\\
5.& Disagree strongly&15.0\%\\
\end{tabular}
\end{enumerate}

\begin{figure}[!hptb]
{\par\centering \resizebox*{10cm}{8cm}{\includegraphics{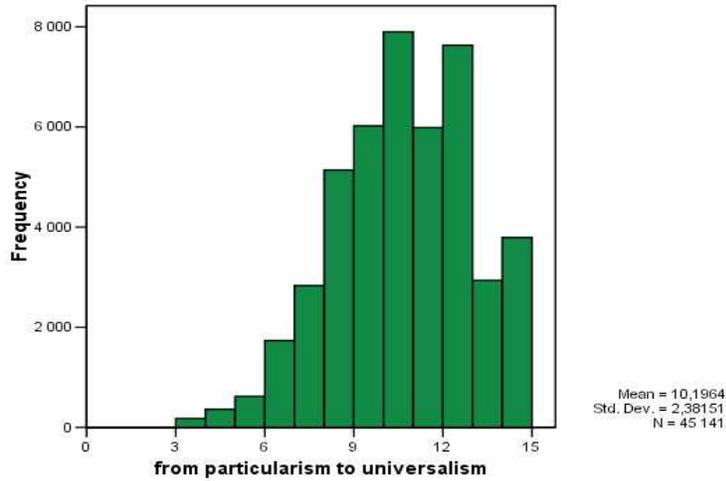}} \par}
\caption{Histogram of index.}
\label{fig-1}
\end{figure}

Based on these variables I made simple index which rage from 3 (the most particularistic attitudes) to 15 (the most universalistic attitudes). Its distribution is shown in Fig. \ref{fig-1}.

On average European respondents are more likely universalist than particularist (the average total mean 10.2).
The ESS was conducted in 24 European countries.
The differences between countries in the mean level of analyzed attitudes range from 9.3 for Belgium (9.5 for Ukraine and 9.7 for Poland --- three most particularistic countries) to 11.1 for Portugal (11.0 for Norway and 10.9 for France --- three most universalistic countries).

%% \begin{table}
%% \label{tab1}
%% \begin{center}
%% \begin{tabular}{|l|c|c|c|}
%% \hline
%% Country&mean&$N$&standard deviation\\
%% \hline
%% Belgium&9.3443&1769&2.69706\\
%% Ukraine&9.5305&1983&2.44040\\
%% Poland&9.7051&1699&2.34700\\
%% Slovenia&9.7419&1422&2.15224\\
%% Austria&9.7753&2194&2.69706\\
%% Netherlands&9.8588&1877&2.17683\\
%% Switzerland&9.8668&2140&2.27237\\
%% Estonia&9.8672&1927&2.31457\\
%% Slovakia&9.8826&1482&2.18422\\
%% Luxemburg&9.8927&1622&2.18109\\
%% Hungary&10.0202&1486&2.31572\\
%% Greece&10.0933&2347&2.23364\\
%% Germany&10.1234&2861&2.31772\\
%% Czech Republic&10.2929&2953&2.57829\\
%% Spain&10.3832&1652&2.24680\\
%% Ireland&10.4370&2270&2.08422\\
%% United Kingdom&10.5539&1892&2.11082\\
%% Sweden&10,5707&1938&2.33321\\
%% Iceland&10.5790&563&2.24450\\
%% Finland&10.6209&2010&2.41605\\
%% Denmark&10.7461&1477&2.39548\\
%% France&10.8520&1804&2.71818\\
%% Norway&11.0222&1758&2.20649\\
%% Portugal&11.0998&2015&2.31494\\
%% Total&10.1964&45141&2.38151\\
%% \hline
%% \end{tabular}
%% \end{center}
%% \end{table}

\begin{figure}[!hptb]
{\par\centering \resizebox*{10cm}{8cm}{\includegraphics{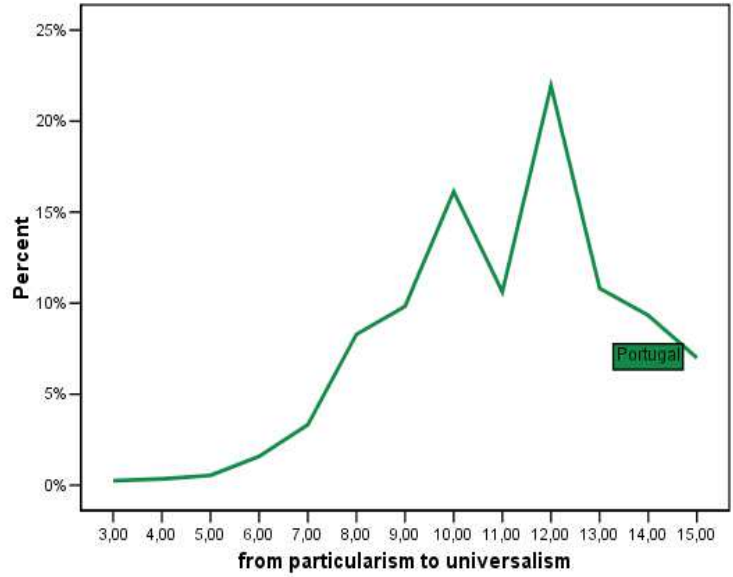}} \par}
\caption{Frequency polygon of Portugal.}
\label{fig-2}
\end{figure}

\begin{figure}[!hptb]
{\par\centering \resizebox*{10cm}{8cm}{\includegraphics{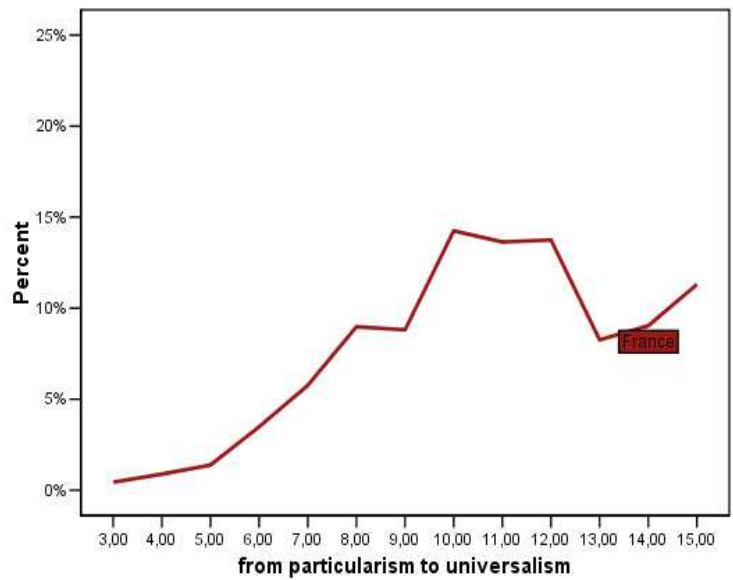}} \par}
\caption{Frequency polygon of France.}
\label{fig-3}
\end{figure}

The results of this simple analysis are mostly coherent with other research findings eg mentioned above Trompenaars and Hampden-Turner conclusions. However, there are two surprising cases --- France and Portugal (see Figs. \ref{fig-2} and \ref{fig-3}). Both are predominantly Catholic countries with managers of particularistic attitudes rather, here with the highest score on universalistic pole of the index --- Portugal and the third highest score on it --- France.  Judging from the shape of a frequency polygon (presented on the figures below) there might be different explanations embedded in social structure for both countries. It will require further analysis to find causes of these discrepancies  but this task goes beyond presented text. 

\section{A case of Poland}
\begin{figure}[!hptb]
{\par\centering \resizebox*{10cm}{8cm}{\includegraphics{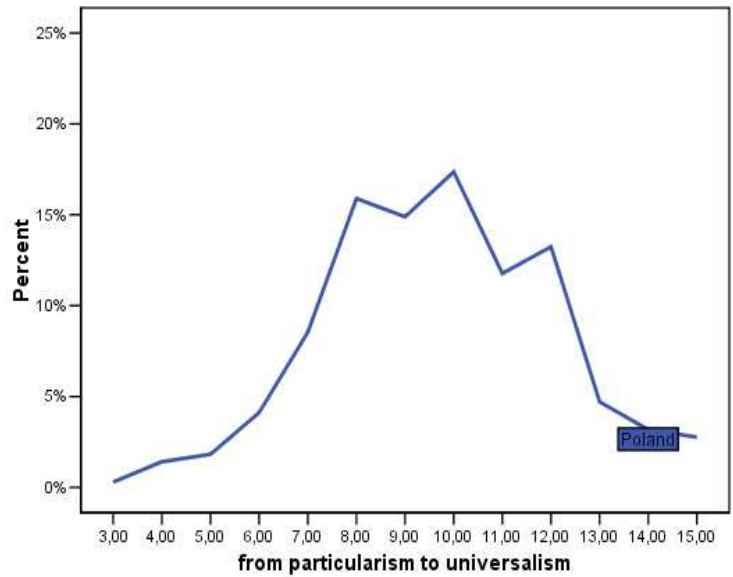}} \par}
\caption{Frequency polygon of Poland.}
\label{fig-4}
\end{figure}

I will be particularly interested in the place of Poland on the mosaic of European cultures of capitalism. Polish culture of capitalism is particularistic rather than universalistic (see Fig. \ref{fig-4}). More specifically I would say that the dilemma between universalism and particularism in Polish culture has been solved in such a way that Poles accept and understand universalistic rules as long as they concern theoretical issues. When it comes to practice they prefer more particularistic solutions. In the everyday life in Poland we can find plenty of examples of disobeying universalistic rules and regulations due to the ``exceptional character'' of cases. This is a more common situation than obeying the law. These attitudes towards legal regulations and formal institutions (e.g. state, company, or contracts) can also be shaped by some historical experiences (e.g. lack of independence during partition time and the communist system). Because of this, formal institutions are still perceived as a part of  ``outside'' world, created from above. Thus regulations are not serving citizens but controlling them \cite{ref-39}. Polish particularism is also described as familiarism \cite{ref-39} which divides the world between a friendly small circle of family and friends and a larger, alien, outside part. This division has consequences in social relations and attitudes toward work.

However, emergence of Polish culture of capitalism will occur, most probably, within the framework of the European Union. This will influence both the direction and the scope of changes in the Polish economy. As the experience of the EU shows, the process of integration does not mean unification. In the enlarged EU, the cultural differences and interests will still have to be managed on the supranational level, as well as on the level of individual companies investing in foreign places. In order to manage these differences successfully, knowledge from the presented kind of research will be very useful.

\end{document}